\documentclass[acmsmall, screen, nonacm]{acmart}
\AtBeginDocument{%
  \providecommand\BibTeX{{%
    \normalfont B\kern-0.5em{\scshape i\kern-0.25em b}\kern-0.8em\TeX}}}

\setcopyright{acmcopyright}
\copyrightyear{2020}
\acmYear{2020}
\acmDOI{10.1145/xxxxx.xxxxx}

\acmConference[Placeholder '20]{Placeholder '20: ACM Symposium on Placeholder}{June 03--05, 2020}{Placeholder, A}
\acmBooktitle{Placeholder '20: ACM Symposium on Placeholder,
  June 03--05, 2020, Placeholder, A}
\acmPrice{0.00}
\acmISBN{xxx-1-4503-XXXX-X/20/06}

\usepackage{booktabs}
\usepackage{tabulary}
\usepackage{tabularx}
\usepackage{longtable}
\usepackage{tabu}
\usepackage{subcaption}
\begin{document}

\title[A Survey on Synchronous AR, VR and MR Rem. Collab. Systems]{A Survey on Synchronous Augmented, Virtual and Mixed Reality Remote Collaboration Systems}


\iffalse
\author{PlaceHolder Name}
\affiliation{%
  \institution{PlaceHolder institution}
  }
\email{placeholder@placeholder.com}

\author{PlaceHolder Name}
\affiliation{%
  \institution{PlaceHolder institution}
  }
\email{placeholder@placeholder.com}

\author{PlaceHolder Name}
\affiliation{%
  \institution{PlaceHolder institution}
  }
\email{placeholder@placeholder.com}
\fi
\author{Alexander Schäfer}
\affiliation{%
  \institution{TU Kaiserslautern}
  }
\email{alexander.schaefer@dfki.de}

\author{Gerd Reis}
\affiliation{%
  \institution{German Research Center for Artificial Intelligence}
  }
\email{gerd.reis@dfki.de}

\author{Didier Stricker}
\affiliation{%
  \institution{German Research Center for Artificial Intelligence, TU Kaiserslautern}
  }
\email{didier.stricker@dfki.de}

\renewcommand{\shortauthors}{Schäfer et al.}
\begin{abstract}
Remote collaboration systems have become increasingly important in today’s society, especially during times where physical distancing is advised. 
Industry, research and individuals face the challenging task of collaborating and networking over long distances. 
While video and teleconferencing are already widespread, collaboration systems in augmented, virtual, and mixed reality are still a niche technology.
%Systems for remote collaboration in virtual or augmented reality are developing slowly but steadily, yet they are not well established.
%Such a system consists of many different facets and requires expertise in various fields such as 3D modeling, animation, development of interactive systems, creation of avatars and dynamic content, multi-user communication and more. 
%In addition, mixed reality systems are often implemented using a combination of AR and VR hardware, which requires a certain expertise in both technologies.
We provide an overview of recent developments of synchronous remote collaboration systems and create a taxonomy by dividing them into three main components that form such systems: \textit{Environment}, \textit{Avatars}, and \textit{Interaction}. 
A thorough overview of existing systems is given, categorising their main contributions in order to help researchers working in different fields by providing concise information about specific topics such as avatars, virtual environment, visualisation styles and interaction.
The focus of this work is clearly on synchronised collaboration from a distance.
A total of 82 unique systems for remote collaboration are discussed, including more than 100 publications and 25 commercial systems.
\end{abstract}




\maketitle

\section{Introduction}

Augmented Reality (AR), Virtual Reality (VR) and Mixed Reality (MR) technologies are becoming more mature and open new ways for remote collaboration.
Video and teleconferencing systems are already in extensive use in today's society, enabling a focus on more novel alternatives which utilize virtual, augmented and mixed reality technology.

Systems for remote collaboration in AR, VR and MR are developing slowly but steadily, yet they are not well established.
Such a system consists of many different facets and requires expertise in various fields such as 3D modeling, animation, development of interactive systems, creation of avatars and dynamic content, multi-user communication and more. 
In addition, mixed reality systems are often implemented using a combination of AR and VR hardware, which requires a certain expertise in both technologies.

In this paper we give a thorough summary and discussion of synchronous remote collaboration systems which utilize AR/VR/MR technology. 
The importance of such systems was emphasized during the COVID-19 outbreak in December 2019.
People all over the world were put into quarantine, cities and local communities forbid traveling and even stepping outside. During this time, the scientific community was forced to find novel ways to network and communicate.
Most scientific conferences where either cancelled or held completely virtual, some even using immersive 3D experiences utilizing VR HMD's.
The IEEE VR 2020 conference as an example, used virtual rooms where conference participants could join and interact with each other.
Paper and poster presentations where done within a virtual environment which was streamed online for a broader audience.
Although video and teleconferencing systems in particular experienced a significantly increased use as a result of this global crisis, the AR, VR and MR community received a major awareness push as well.

A well sophisticated  AR/VR/MR system could help to reduce travel costs, office space, time, and carbon emissions by creating shared immersive spaces with believable person embodiment and interaction. To compete with each other in this crisis, many companies and researchers have recently invested in creating novel systems, which makes a recent review of existing systems and research even more interesting.
\par
We identify and classify the individual parts which are necessary for a remote collaboration system and provide an overview of existing systems for research and professional work. Remote collaboration systems utilizing AR/VR/MR technology are used in many different fields such as human computer interaction, computer graphics, medicine, training, cognitive sciences and many more. We define three components that each remote collaboration system needs to implement, namely \textit{Environment}, \textit{Avatars} and \textit{Interaction}. A detailed explanation about this taxonomy is described in section \ref{sec:pillars}. 
By providing condensed information on certain key topics, we want to help researchers assess the state of the art in a particular subject. As an example, a researcher focused on interaction in multi-user collaborative environments will be intersted in inspecting Table \ref{tab:interaction} which categorises  important works in regard to interaction types which where found during the survey.
A researcher focused on novel environments for remote collaboration can use Table \ref{tab:environment2} which lists the discussed work with respect to their technology, use case, visualisation style and the stimulated sensory inputs. The representation of other users, in regard to their visualisation and animation style is shown in Table \ref{tab:avatar}.

\section{Related Work and Survey Procedure}
\label{sec:relatedwork}
Related work was conducted by Phon et al. \cite{phon2014collaborative} which reviewed the state of the art in collaborative AR systems with focus on education in 2014. This work has a clear focus on collaborative learning in AR and does not differentiate between remote and local collaboration experiences. Another survey was done by Wang et al \cite{wang2019comprehensive} with focus on AR and MR based collaborative design in manufacturing. 
Ens et al. \cite{ens2019revisiting} review published work in collaboration through mixed reality up to the year 2018. Although the focus of mentioned work is not remote collaboration explicitely, the authors differentiate between remote and physically co-located systems. Another Survey was conducted by de Belen et al. \cite{de2019systematic} in which the authors provide a systematic review of collaborative mixed reality technologies.\par

\subsection{Inclusion and Exclusion of Systems}
\label{sec:inclusion}
In our work we provide a concise focus on \textit{synchronous remote collaboration systems} and we categorise the results to assist scientists in different fields to cover their specific research interest. We emphasize the term \textbf{remote}, which means that physical co-location of users is not required and the term \textbf{synchronous} which allows users to collaborate in real-time. The focus is on virtual, augmented and mixed reality systems. Traditional video and teleconferencing systems are omitted. We define a remote collaboration system as a way of communicating, interacting and sharing a space beyond the boundaries of physical space exclusively through technological channels with distributed users. We include work which uses or implements a combination of AR/VR/MR technology and synchronous remote collaboration. Systems which allow multiple users in a system but require users to be in physical co-location are excluded. Exceptions are systems which where used in a physical co-location scenario but could easily be extended for remote collaboration purposes. Asynchronous systems that do not allow real-time communication between users are also excluded. 

\subsection{Methodology}
\label{sec:methodology}
The survey was conducted through an iterative process by integrating the most relevant papers first, identifying specific similarities and differences with subsequent categorisation. 
By incrementally adding new relevant research work we evolved the categorisation process and therefore separated the relevant work into three main contribution categories: \textit{Environment}, \textit{Interaction} and \textit{Avatars}. 
With this approach it is possible to filter the relevant papers (over 2.000 unique papers) by applying the constraints mentioned in section \ref{sec:inclusion} and then fitting the remaining papers into categories of the specific research interest for researchers.

\begin{figure}[H]
\centering
\includegraphics[width=1.0\columnwidth]{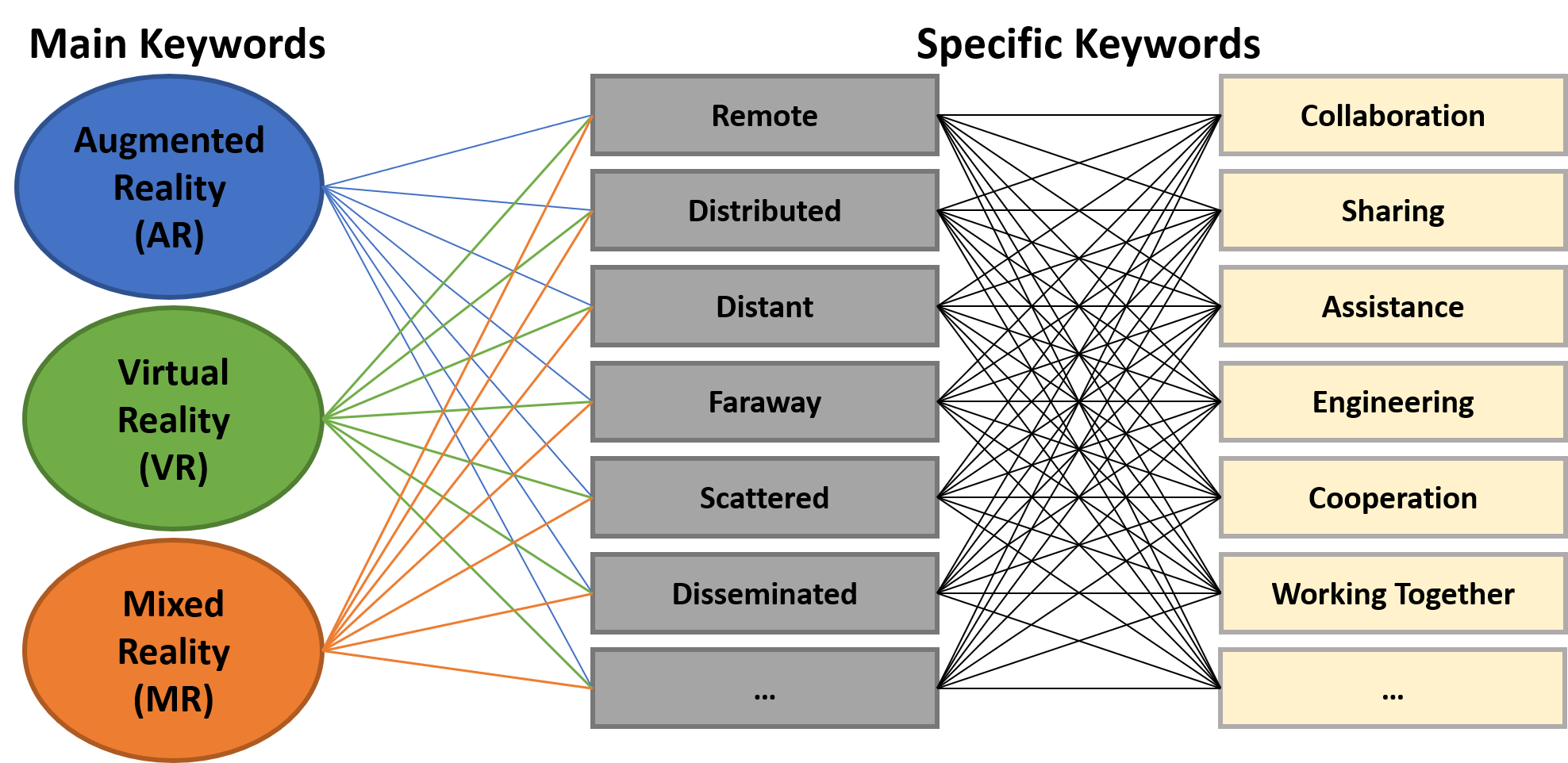}
\caption{The used search methodology.}
\label{fig:searchmethod}
\end{figure}

Our extensive search was performed by using search queries in different data sources. The used data sources include Scopus (\url{https://www.scopus.com}), Google Scholar (\url{https://scholar.google.com/}), ACM Digital Library (\url{https://dl.acm.org/}), IEEE Xplore (https://ieeexplore.ieee.org/), Springer Link (\url{https://link.springer.com/}), PubMed (\url{https://pubmed.ncbi.nlm.nih.gov/}) and Microsoft Academic (\url{https://academic.microsoft.com/}).
Additionally we looked through the proceedings of multiple leading AR and VR conferences such as IEEE VR, ISMAR and EuroVR with focus on collaboration related topics.
The search was performed by concatenating AR/VR/MR with keywords such as \textit{remote}, \textit{collaboration}, \textit{social} and more (as shown in Figure \ref{fig:searchmethod}).
The tables in the following chapters summarise works from the same author if it is a continuation or extension of the previous work. Furthermore, the focus is on the general implementation of the proposed systems rather than on their specific research questions.

\subsection{Definition AR, VR and MR systems}
\begin{figure}[H]
\centering
\includegraphics[width=1.0\columnwidth]{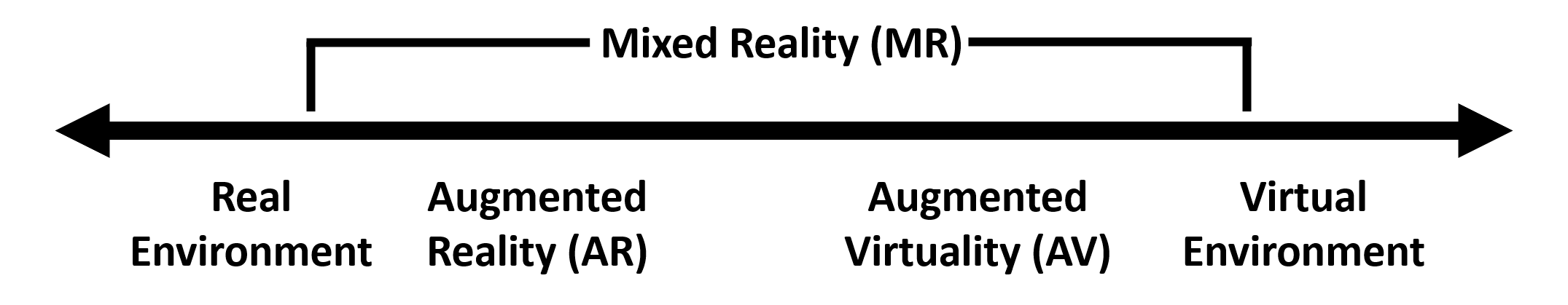}
\caption{The Reality-Virtuality continuum according to Milgram et al. \cite{milgram1995augmented}.}
\label{fig:VC}
\end{figure}

In this section we briefly introduce how we perceive AR, VR and MR systems.
In general, AR is achieved with two approaches: \textit{Video-Seethrough} and \textit{Optical-Seethrough}.
In both cases, the real world is augmented to the user. 
In Video-Seethrough the world is captured with a camera and virtual objects are placed onto the captured images. 
In Optical-Seethrough systems, users perceive the outside world with their own eyes through a transparent projection surface which displays the AR content. 
Regarding VR, most literature couples the term VR with a Head Mounted Display (HMD) which is placed on the head of a user. 
In this survey we also consider systems without HMD's as VR, independent of the specific display device, as long as it is possible to immerse users into a virtual 3D environment.
While AR and VR systems are often quite clear in their separation, the distinction between MR systems often leads to confusion. 
For this survey, we use Milgram et al.'s \cite{milgram1995augmented} definition of the Reality-Virtuality continuum which describes mixed reality as the area where both, the real and virtual world are mixed (see Figure \ref{fig:VC}).

More precisely, a system is referred to as MR in this survey if at least one of the following points applies:
\begin{enumerate}
    \item There is a mix betweeen AR and VR hardware (this also includes projector based systems)
    \item Real world objects are used for interaction with either AR or VR hardware
\end{enumerate}
Although the recently coined term Extended Reality XR is very popular, we do not use it in this context because we explicitly exclude "pure reality" and the remaining parts are covered by the terms AR, VR and MR.

\subsection{The three pillars for remote collaboration systems: A taxonomy of remote collaboration systems}
\label{sec:pillars}
\begin{figure}[H]
\centering
\includegraphics[width=0.5\columnwidth]{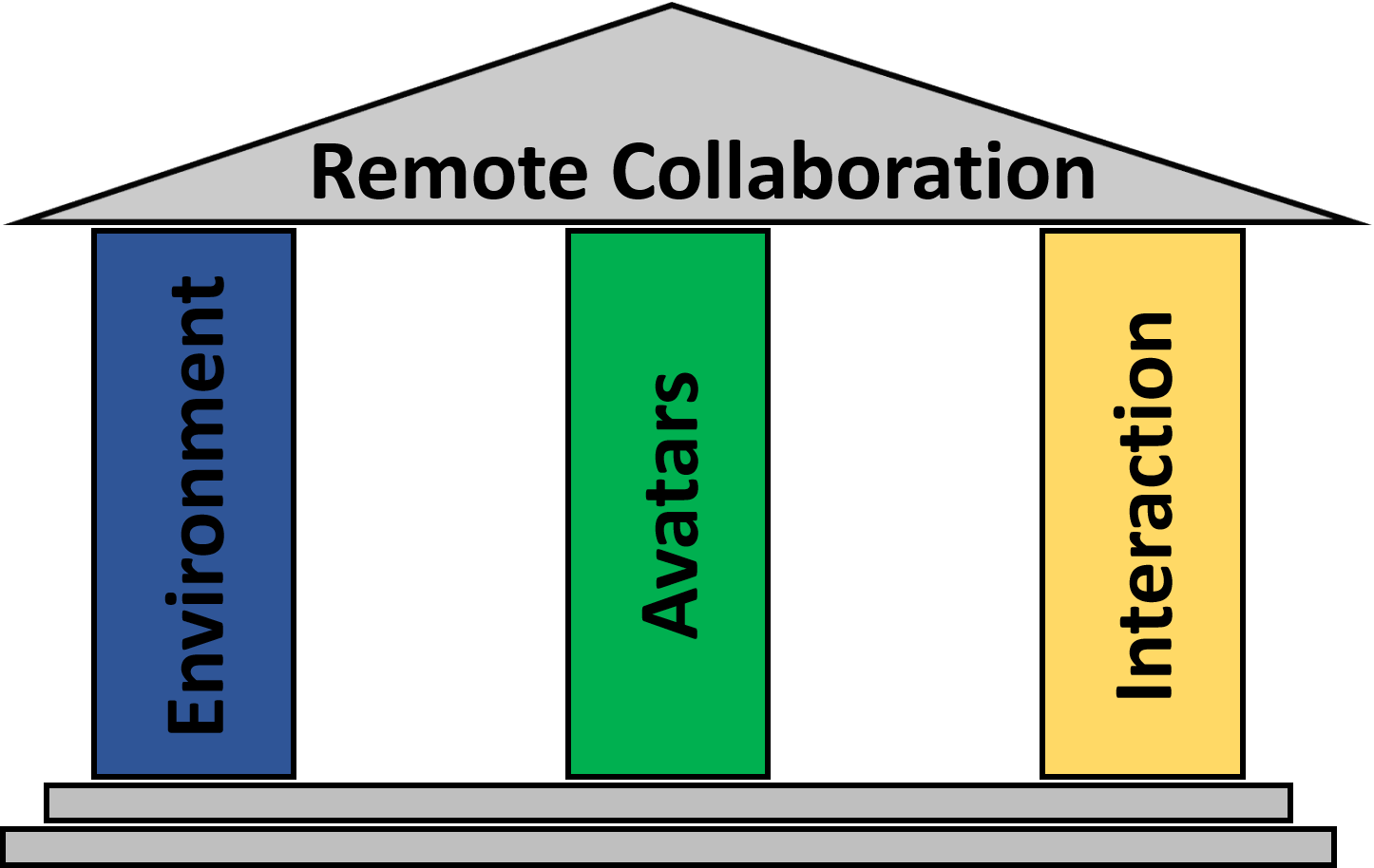}
\caption{The three pillars for remote collaboration systems.}
\label{fig:threepillars}
\end{figure}

We create a taxonomy and categorise the relevant work in a logical manner.
Many systems have different aspects of novelty which cannot be described by assigning them to a specific category. 
E.g. one system might excel in the novelty of avatars while another introduced a new kind of interaction technique for remote collaboration.
One goal of this survey is to help researchers from a wide range of fields who are interested in the area of remote collaboration systems which utilize AR, VR and MR technology.
To illustrate this: A researcher who is interested in the topic remote collaboration using AR/VR/MR might ask "How are users represented in virtual environments?", "What kind of interaction is possible in a shared virtual space?" or "Are there collaboration systems which enable shared gaze awareness?".
With our survey we want to provide condensed information to different research questions such as virtual representation of users, different types of interaction and the virtual environment.
To achieve this, we elaborated a concept that enables the possibility to view each of these systems from different viewpoints: \textit{Environment}, \textit{Avatars} and \textit{Interaction} (see Figure \ref{fig:threepillars}) which we call \textbf{the three pillars of remote collaboration systems}.
In the next chapters we explain each of these components more detailed and present important and highly cited publications in each category. Additionally we provide tables for each category to allow quick access to the desired work: Table \ref{tab:environment2} is summarizing remote collaboration systems with focus on virtual environment, Table \ref{tab:avatar} focuses on user representation and Table \ref{tab:interaction} identifies and categorises different interaction possibilities.

\section{Components of Remote Collaboration Systems}

\subsection{Environment}
\label{sec:environment}

\begin{table*}
\centering
\setlength\tabcolsep{1.5pt} % default value: 6pt
\begin{tabulary}{\textwidth}{LCCCL}
\toprule
System / Authors & \hfil Techn. & \hfil Use Case  & \hfil Visualisation Style & \hfil Sensory Inputs \hfil\\
%\midrule
\cmidrule(rl){1-1}\cmidrule(rl){2-2}\cmidrule(rl){3-3}\cmidrule(rl){4-4} \cmidrule(rl){5-5} 
% VR %%%%%%%%%%%%%%%%%%%%%%%%%%%%%%
Breakroom \cite{breakroom} & VR & Meeting &  Cartoon & Audio, Visual \\
EngageVR \cite{EngageVR}& VR& Meeting &  Realistic & Audio, Visual\\   
Glue Collab \cite{Glue}& VR& Meeting &  Realistic & Audio, Visual\\   
MeetInVR \cite{MeetInVR}& VR&  Meeting  &  Realistic & Audio, Visual\\   
Mozilla Hubs \cite{MozillaHubs}& VR&  Meeting & Cartoon & Audio, Visual \\   
Nvidia Holodeck \cite{NvidiaHolodeck}& VR &  Meeting &  Realistic & Audio, Visual \\   
Stage VR \cite{StageVR}& VR &Meeting  &  Realistic & Audio, Visual\\
TechViz VR \cite{TechViz}& VR&  Meeting  &  Realistic & Audio, Visual \\
TheWild \cite{TheWild}& VR & Meeting &  Realistic & Audio, Visual\\
Vive Sync \cite{ViveSync}& VR& Meeting &  Cartoon & Audio, Visual\\
WorldViz \cite{WorldViz}& VR&  Meeting & Cartoon & Audio, Visual\\
Regenbrecht et al. \cite{regenbrecht2004using}& VR & Meeting& Realistic & Audio, Visual \\
Gu et al. \cite{gu2011technological}& VR & Meeting & Realistic & Audio, Visual \\
Schäfer et al. \cite{schafer2019towards} & VR & Meeting & Realistic & Audio, Visual \\
VRChat \cite{chat2018create} & VR & Meeting & Cartoon & Audio, Visual \\
NeosVR \cite{NeosVR} & VR & Meeting & Cartoon & Audio, Visual \\
Acadicus \cite{Acadicus} & VR & Meeting & Cartoon & Audio, Visual \\
Rumii \cite{Rumii} & VR & Meeting & Realistic & Audio, Visual \\
VirBELA \cite{VirBELA} & VR & Meeting & Realistic & Audio, Visual \\
Garou \cite{Garou} & VR & Meeting & Realistic & Audio, Visual \\
MeetingRoom \cite{MeetingRoom} & VR & Meeting & Cartoon & Audio, Visual \\
Facebook Horizon \cite{Horizon} & VR & Meeting & Cartoon & Audio, Visual \\
Second Life \cite{SecondLife} & VR & Meeting & Realistic & Audio, Visual \\
Tan et al. \cite{tan2017virtual} & VR & Meeting & Cartoon & Audio, Visual\\
Weissker et al. \cite{weissker2020getting} & VR & Meeting & Cartoon & Audio, Visual\\

IrisVR \cite{IrisVR} & VR & Design & Cartoon & Audio, Visual \\
Hsu et al. \cite{hsudesign} & VR & Design & Realistic & Audio, Visual \\
Lehner et al. \cite{lehner1997distributed} & VR & Design & Realistic & Audio, Visual\\
BigScreen \cite{BigScreen} & VR & Event & Realistic & Audio, Visual \\
Wave \cite{Wave} & VR & Event & Realistic & Audio, Visual \\
Sansar \cite{Sansar} & VR & Event & Realistic & Audio, Visual \\
% AR %%%%%%%%%%%%%%%%%%%%%%%%%%%%%%%
Orts et al. \cite{orts2016holoportation} & AR & Meeting & Realistic & Audio, Visual \\
Regenbrecht et al. \cite{regenbrecht2002magicmeeting} & AR & Meeting & Realistic & Audio, Visual \\
Shen et al. \cite{shen2010augmented, shen2006framework, shen2008product, shen2008collaborative} & AR & Design & Cartoon & Audio, Visual, Tactile \\
Poppe et al. \cite{poppe2011prototype, poppe2012preliminary} & AR & Design & Cartoon & Audio, Visual \\
Sodhi et al. \cite{sodhi2013bethere} & AR & Remote Expert & Cartoon & Visual \\
Gurevich et al. \cite{gurevich2015design, gurevich2012teleadvisor} & AR & Remote Expert & Annotations & Audio, Visual \\
Masai and Lee et al. \cite{masai2016empathy, lee2016remote} & AR & Remote Expert & Cartoon & Audio, Visual \\
Tait et al. \cite{tait2015effect} & AR & Remote Expert & Cartoon & Audio, Visual \\
Kurata et al. \cite{kurata2004remote} & AR & Remote Expert & Cartoon & Audio, Visual \\
Ou et al. \cite{ou2003dove} & AR & Remote Expert & Cartoon & Audio, Visual \\
Izadi et al. \cite{izadi2007c} & AR & Remote Expert & Cartoon & Audio, Visual \\
Lukosch et al. \cite{lukosch2012novel} & AR & Remote Expert & Annotations & Audio, Visual \\
    \bottomrule
\end{tabulary}
\label{tab:environment}
\end{table*}

\begin{table*}
\centering
\setlength\tabcolsep{1.5pt} % default value: 6pt
\begin{tabulary}{\textwidth}{LCCCL}
\toprule
System / Authors & \hfil Techn. & \hfil Use Case  & \hfil Visualisation Style & \hfil Sensory Inputs \hfil\\
%\midrule
\cmidrule(rl){1-1}\cmidrule(rl){2-2}\cmidrule(rl){3-3}\cmidrule(rl){4-4} \cmidrule(rl){5-5} 
% AR %%%%%%%%%%%%%%%%%%%%%%%%%%%%%%%
Gauglitz et al. \cite{gauglitz2014touch, gauglitz2012integrating, gauglitz2014world} & AR & Remote Expert & Annotations & Audio, Visual \\
Gupta et al. \cite{gupta2016you} & AR & Remote Expert & Annotations & Audio, Visual \\
Zillner et al. \cite{zillner2018augmented} & AR & Remote Expert & Annotations & Visual \\
Utzig et al. \cite{utzig2019augmented} & AR & Remote Expert & Annotations & Audio, Visual \\
Zenati et al. \cite{zenati2013new, zenati2015maintenance, zenati2014augmented} & AR & Remote Expert & Annotations & Audio, Visual \\
% MR %%%%%%%%%%%%%%%%
Higuchi et al. \cite{higuchi2015immerseboard} & MR & Meeting & Realistic & Audio, Visual \\
Speicher et al. \cite{speicher2018360anywhere} & MR  & Meeting & Annotations & Audio, Visual \\
Ryskeldiev et al. \cite{ryskeldiev2018spatial, ryskeldiev2018streamspace} & MR  & Meeting & Annotations & Visual \\
Spatial \cite{Spatial} & MR &  Meeting &  Realistic & Audio, Visual\\   
Regenbrecht et al. \cite{regenbrecht2006carpeno} & MR  & Meeting & Realistic & Audio, Visual, Tactile \\
Haller et al. \cite{haller2005coeno} & MR  & Meeting & Realistic & Audio, Visual, Tactile \\
Norman et al. \cite{norman2019impact} & MR & Meeting & Annotations & Audio, Visual \\
Matthes et al.  \cite{matthes2019collaborative} & MR & Meeting & Realistic & Audio, Visual \\
Galambos et al. \cite{galambos2015design, galambos2014merged} & MR & Meeting & Realistic & Audio, Visual \\
Bai et al. \cite{bai2020user} & MR & Meeting & Cartoon & Audio, Visual \\
Luxenburger et al. \cite{luxenburger2016medicalvr} & MR & Meeting & Realistic & Audio, Visual\\
vTime \cite{VTime} & MR & Meeting & Realistic & Audio, Visual \\
PoseMMR \cite{pan2020posemmr} & MR & Meeting & Annotations & Audio, Visual \\
Gr{\o}nb{\ae}k et al. \cite{gronbaek2001interactive} & MR  & Design & Realistic & Audio, Visual \\
TeleAR \cite{wang2014mutual}  & MR & Design & Cartoon & Audio, Visual, Tactile \\
Wang et al. \cite{wang2007exploring, wang2009experimental, wang2006potential, wang2008user} & MR  & Design & Realistic & Audio, Visual, Tactile \\
Sakong et al. \cite{sakong2006supporting} & MR  & Design & Realistic & Audio, Visual, Tactile \\
Sidharta et al. \cite{sidharta2006augmented} & MR  & Design & Realistic & Audio, Visual, Tactile \\
Ibayashi et al. \cite{ibayashi2015dollhouse} & MR & Design & Realistic & Audio, Visual, Tactile \\
Sasikumar et al. \cite{sasikumar2019wearable} & MR & Remote Expert & Cartoon & Audio, Visual \\
Lee et al. \cite{lee2017sharedsphere, lee2018user} & MR & Remote Expert & Realistic & Audio, Visual \\
Teo et al. \cite{teo2019360drops, teo2019mixed, teo2019technique, teo2019merging} & MR & Remote Expert & Realistic & Audio, Visual \\
Kim et al. \cite{kim2018effect, kim2019evaluating} & MR & Remote Expert & Annotations & Audio, Visual \\
Rae et al. \cite{rae2014bodies} & MR & Remote Expert & Realistic & Audio, Visual \\
Piumsomboon et al. \cite{piumsomboon2017empathic, piumsomboon2017covar} & MR  & Remote Expert& Cartoon & Visual \\
Gao et al. \cite{gao2016oriented, gao2017static} & MR  & Remote Expert& Realistic & Audio, Visual \\
Wang et al. \cite{wang2019head} & MR & Remote Expert & Annotations & Audio, Visual \\

Elvezio et al. \cite{elvezio2017remote} & MR & Remote Expert & Realistic & Visual \\
Pouliquen-Lardy et al. \cite{pouliquen2016remote}  & MR & Remote Expert & Realistic & Audio, Visual \\
Alem et al. \cite{alem2011handsonvideo} & MR & Remote Expert & Cartoon & Audio, Visual \\
Higuch et al. \cite{higuch2016can} & MR & Remote Expert & Realistic & Audio, Visual \\
Chen et al. \cite{chen20153d} & MR & Remote Expert & Annotations & Audio, Visual \\
Nittala et al. \cite{nittala2015planwell} & MR & Remote Expert & Annotations & Audio, Visual, Tactile \\
Sun et al. \cite{sun2018employing, sun2016optobridge} & MR & Remote Expert & Cartoon & Audio, Visual \\
    \bottomrule
\end{tabulary}
\caption{Remote collaboration systems sorted by their respective technology and classified in different categories.}
\label{tab:environment2}
\end{table*}

\textit{Virtual environment} refers to a simulated environment that stimulates the sensory impressions of a user. 
One of the first virtual environments was \textit{Sensorama}, created by Morton Heilig in the early sixties \cite{heilig1962sensorama}. 
It featured a simulated motorcycle ride with 3D visuals, stereo sound, olfactory cues (aromas) and tactile cues (seat vibration and wind from fans).
In recent literature, most virtual environments are not as comprehensive and complete as the prototype created by Heilig, but rather focus on specific areas that are mostly visual or acoustic stimuli.
Exceptions are augmented reality systems which utilize markers, where tangible interfaces with haptic feedback are still popular. As an example, Wang et al. used tangible interfaces such as a regular table \cite{wang2009experimental} or tabletop \cite{wang2008user, wang2014mutual}. 
Other marker based systems used turntables, such as Shen et al. \cite{shen2006framework, shen2008collaborative} or additional interaction tools such as a pen in \cite{shen2008product}. 
With increasing maturity of AR technology, marker based systems became obsolete and such systems are not further developed. \par

In case of VR, there is usually a 3D modeled scene which is rendered, while in AR, the virtual environment refers to the augmented virtual objects superimposed onto the real world. 
Some AR systems do not include any 3D object rendering but use shared annotations and virtual pointers instead \cite{ryskeldiev2018spatial, ryskeldiev2018streamspace, lukosch2012novel, gupta2016you, gurevich2015design, gurevich2012teleadvisor, gauglitz2012integrating, gauglitz2014world, gauglitz2014touch}.
Sense of presence, often called telepresence, is highly affected by the quality and consistency of the virtual environment \cite{yoon2019effect}. 
In early work, studies suggested that the overall sense of presence is increased by adding tactile and auditory cues \cite{dinh1999evaluating}.
The more sensory impressions are added, the greater the feeling of presence according to the studies of Dinh et al. \cite{dinh1999evaluating}.
In VR as example, telepresence is not only achieved by highly realistic 3D environments but also with consistency i.e. avatars should blend in with the environment and interaction methods should be adequate \cite{yoon2019effect}.
The work of Yoon et al. \cite{yoon2019effect} compares different types of avatars with different styles.
The authors' findings include that a virtual environment in cartoon style should also use avatars in cartoon style to achieve a higher sense of presence for the users.
In Table \ref{tab:environment2} we present research works, summarised by their respective virtual environment properties and ordered according to their respective technology (AR/VR/MR). 
Furthermore, we categorised remote collaboration systems in three main use cases: \textit{Meeting, Design} and \textit{Remote Expert} since they where most popular and consistent throughout the literature. The category \textit{Meeting} can also be seen as a means of \textit{sharing a workspace with other users}.
Some systems which are used for training or socializing fit also in this category.
Note that the category \textit{Event} has been added for VR-based systems, as there were three systems that could not otherwise be meaningfully categorised.
These systems are used solely for event purposes. 
Bigscreen \citep{BigScreen} focuses on virtual cinemas, allowing people to buy tickets and then watch movies together in a collaborative virtual environment. Sansar \cite{Sansar} and Wave \cite{Wave} focus on virtual live events such as concerts. To the best of our knowledge, there is no existing system purely designed for events based on AR/MR technology.

\subsubsection{Category Meeting}
This category is for remote collaboration systems where users share a common workspace or environment for collaboration.
These systems usually support media sharing, involve avatars to increase the sense of co-presence, and have interactive elements such as drawing on a whiteboard. In addition, we include use cases with knowledge transfer i.e. educational and learning scenarios in this category. \par
A focus on transferring and obtaining knowledge through augmented and virtual reality remote collaboration systems is shown by Monahan et al. \cite{monahan2008virtual}, where a web-based VR system for managing and providing educattional content online was implemented. 
The system features an immersive 3D environment, allowing the lecturer to add media and virtual objects.
Avatars are able to use gestures e.g. they can raise hands to indicate a question.

Chen et al. \cite{chen2019immertai} created \textit{ImmerTai}, a system which is designed for remote motion training.
The participants are able to learn Chinese Taichi in an immersive collaborative environment.
Student and a teacher are physically separated and resembled as a full body avatar in the virtual environment.
This system includes a motion capturing module utilizing a Microsoft Kinect, transferring the real world motion to their avatars.
A motion assessment module is used to rate the movements of the student and give hints for improvement during and after a Taichi session.

Wang et al. \cite{wang20192} use a combination of camera, projector, VR HMD and hand tracker to create a remote collaboration system for knowledge transfer in a manufacturing scenario.
A local worker assembles a water pump while a camera is recording and transmitting video footage of the worker's assembling progress to a remote expert.
The remote expert views the video material through a VR HMD and transmits visual cues back to the local worker.
A projector on the side of the local worker projects the hand movements of the remote expert onto his working surface. \par
Schäfer et al. \cite{schafer2019towards} used panorama images to create a shared photorealistic virtual environment in a meeting scenario.
Users are able to hold virtual presentations with media sharing and hand gestures for interacting with the augmented virtual objects. \par
Weissker et al. \cite{weissker2020getting} investigated group navigation in virtual immersive environments.
The authors implemented a system which allowed users to navigate inside virtual environments together as a group or as individual. In their system, users can attach themselves to others and then organize teleportation movement through the virtual world together.
Their results showed advantages in collaborative work when a switch between individual and group navigation is implemented.

\subsubsection{Category Design}
The categery design combines remote collaboration systems such as product design \cite{shen2010augmented, shen2006framework, shen2008product} and architectural design \cite{hsudesign, chowdhury2019laypeople, hong2019architectural}.\par
One of the earlier works was done by Lehner and DeFanti \cite{lehner1997distributed}, who used a CAVE system in 1997. CAVE is a 10-foot by 10-foot by 9-foot surround screen which uses projections on the walls and floor.
The authors implemented a system which enabled multiple users to share the same environment and discuss vehicle design remotely. 
The visual representation of other users was achieved by streaming 2D video inside the virtual environment.\par
Hsu et al. \cite{hsudesign} developed an architectural design discussion system with interactive and immersive elements. 
It features voice communcation, object manipulation, mid-air sketching and on-surface sketching.
Overall, this tool was implemented to help architects to better understand architecture modeling and to discuss design decisions, even changing models during a remote collaboration session.

The work of Chowdhury et al. \cite{chowdhury2019laypeople} implements a collaboration system with an immersive virtual environment specifically created for urban design ideation and generation.
The work is intended to be used by non-experts and concludes that even laypeople can take part in the design process of early stage urban design. 
While one user uses a VR HMD to view, interact and change virtual objects, other participants are able to perceive the changes on a display screen while giving feedback to the VR user. 

Hong et al. \cite{hong2019architectural} utilizes the multi-user virtual environment Second Life \cite{SecondLife} for creative collaboration with focus on architectural design.
They compared the effectiveness of collaborative architectural design between multi user virtual environments and a commercial architectural design software which allows 2D sketching and communication through audio.
The authors argue that a remote collaboration system with avatars is more effective than two-dimensional approaches due to shared spatial informations.

Ibayashi et al. \cite{ibayashi2015dollhouse} created a MR collaboration system which connects users on a tabletop device with a user wearing a VR HMD.
The authors use a role based system with designers and occupants.
Designers are able to view a 3D environment using a tabletop device. The environment can be changed with a touch interface provided to the designers.
The occupant is immersed in this shared 3D environment with a VR HMD and is able to see the changes made by the designers in real-time.
A see-through ceiling allows the occupant to see the designers by looking at the ceiling of the 3D environment while the designers are able to see the VR user moving around from the top-view. \par
A petroleum well planning application was developed by Nittala et al. \cite{nittala2015planwell}, using hand held devices to augment the surroundings of a remote worker.
A local user used a 3D printout, stylus and tablet as an interface to communicate with a remote worker who is on-site coordinating drilling operations.
The 3D printout was combined with AR visualisations to provide the local user with an overview of earth's composition near the remote worker.
The remote worker is able to see AR annotations made by the local user to plan drilling operations.

\subsubsection{Category Remote Expert}
We identified several systems which include remote collaboration, and use a scenario with a local and a remote user.
In such systems, the local user typically executes a predefined task, being physically present at the target location, while the remote user is generally far away and provides support with instructions or hints.
In this type of collaboration scenario, the remote user is often called the remote expert.
Many systems are based on a combination of AR and VR \cite{lee2017sharedsphere, teo2019technique, gauglitz2014world, gao2017real}, while the remote expert typically uses a VR HMD or 2D screen and the local user transmits his surroundings with the help of an AR HMD or a mounted camera. \par
A mixed realiy collaboration system was developed by Piumsomboon et al.\cite{piumsomboon2017exploring}. 
The system enables an AR user to share his local environment with a remote user. 
It provides collaborative, natural interaction with gaze and hand gesture data transmitted over a network to each user.
 
Another MR collaboration system was developed by Lee et al. \cite{lee2017sharedsphere}. 
The authors developed a system in which a host works with an AR HMD mounted with a 360° camera and a guest with a VR HMD.
Nonverbal communication cues are transmitted via hand tracking and view awareness.
Both users have visual feedback where the counterpart is currently looking at and are able to exchange hand gestures.

Teo et al. \cite{teo2019technique} introduced a MR remote collaboration system which combines reconstructed scenes obtained through an AR HMD with 360 panorama images. 
A remote user who receives footage from the AR user can move through the transmitted visual information, without relying on the local user to move. The system has been extended with the functionality that the remote user can trigger a 360 camera with the help of his VR controller to save spherical images that can be accessed independently \cite{teo2019merging}. The authors propose to add more functionality such as mid-air drawing in 3D to improve the usability of the prototype.  \par

Mixed reality remote collaboration systems supporting local and remote users are especially useful in repairing tasks as Gauglitz et al. \cite{gauglitz2014world} suggest. 
In their work, a remote user is able to see the local user's current view and to annotate the view which is then visible in AR.

An interesting remote collaboration approach is \textit{360Anwywher}e, a framework by Speicher et al \cite{speicher2018360anywhere}. 
It allows ad-hoc remote collaboration in AR via 360-degree live input. 
Users are able to add digital annotations by drawing on a 360-degree video stream, either by means of a normal desktop application or mobile devices.
The annotations made by remote participants are then visualised at the local physical space through a projector.

Gao et al. \cite{gao2017real} implements a mixed reality collaboration system by mounting an RGB and RGB-Depth camera on top of a local user's VR HMD. 
The VR HMD is used as a Video-Seethrough device while it captures and transmits its view to the remote counterpart.
A RGB-Depth camera is used to obtain a point cloud which is streamed to the remote user. 
This point cloud is then stitched, enabling an independent view control of the local workspace. \par

Bai et al. \cite{bai2020user} developed a system which supports real-time 3D reconstruction by assembling eight RGB-Depth cameras into one sensor cluster. 
The remote VR user is able to see the local users surroundings through the transmission of the aligned pointclouds obtained through the RBG-Depth cameras at the local users space.
The authors research focus is a shared virtual environment which supports gaze and gesture as visual cues in a remote expert scenario.
Although the system supports one-way transmission of natural communication cues only,  the results demonstrate advantages by providing natural gaze and gesture cues during collaboration. \par

Overall, we found that the virtual environment of remote collaboration systems focus on audiovisual stimuli.
Some work did not even implement audio, focusing completely on visual feedback \cite{ryskeldiev2018spatial, ryskeldiev2018streamspace, piumsomboon2017covar, piumsomboon2017empathic}.
While other work included some tactile feedback, these systems where mostly marker based AR systems and rely markers attached to physical objects \cite{sidharta2006augmented, sakong2006supporting, wang2008user, wang2007exploring}.

\subsection{Avatars}
\label{sec:avatars}
\begin{table}[t]
\centering
\begin{tabulary}{0.95\textwidth}{p{10em}p{28em}}
\toprule
\textbf{Avatar Type} & \hfil \textbf{References}\\
%\midrule
\cmidrule(rl){1-1}\cmidrule(rl){2-2}
Cartoon& \cite{breakroom} \cite{Glue} \cite{MeetInVR} \cite{MozillaHubs} \cite{StageVR} \cite{ViveSync} \cite{poppe2011prototype, poppe2012preliminary} \cite{schafer2019towards} \cite{NeosVR} \cite{IrisVR} \cite{Acadicus} \cite{Rumii} \cite{RecRoom} \cite{VirBELA} \cite{Garou} \cite{MeetingRoom} \cite{BigScreen} \cite{Wave} \cite{Horizon} \cite{VTime} \cite{teo2019merging} \cite{sasikumar2019wearable} \cite{chat2018create} \cite{galambos2015design, galambos2014merged} \cite{bai2020user} \cite{weissker2020getting} \cite{pouliquen2016remote} \cite{alem2011handsonvideo} \cite{utzig2019augmented}\\
Realistic & \cite{EngageVR} \cite{NvidiaHolodeck} \cite{Spatial} \cite{TechViz} \cite{orts2016holoportation} \cite{higuchi2015immerseboard} \cite{Sansar} \cite{SecondLife} \cite{tan2017virtual} \cite{chat2018create} \cite{NeosVR} \\   
Full Body & \cite{breakroom} \cite{EngageVR} \cite{NvidiaHolodeck} \cite{StageVR} \cite{ViveSync} \cite{chat2018create} \cite{NeosVR} \cite{Sansar} \cite{Wave} \cite{VTime} \cite{SecondLife} \cite{tan2017virtual}\\   
Head \& Hands& \cite{Glue} \cite{MeetInVR} \cite{TechViz} \cite{WorldViz} \cite{piumsomboon2017covar}  \cite{schafer2019towards} \cite{IrisVR} \cite{Acadicus} \cite{Rumii} \cite{MeetingRoom} \cite{BigScreen} \cite{teo2019merging} \cite{galambos2015design, galambos2014merged} \cite{bai2020user} \cite{pouliquen2016remote} \\   
Upper Body& \cite{MozillaHubs} \cite{Spatial} \cite{TheWild} \cite{poppe2011prototype, poppe2012preliminary} \cite{RecRoom} \cite{Garou} \cite{Horizon} \cite{weissker2020getting}\\   
Reconstructed Model & \cite{EngageVR} \cite{Spatial} \cite{TechViz} \cite{orts2016holoportation}\\   
Video & \cite{regenbrecht2006carpeno} \cite{regenbrecht2004using} \cite{gu2011technological} \cite{ryskeldiev2018spatial, ryskeldiev2018streamspace} \cite{matthes2019collaborative} \cite{lehner1997distributed} \cite{chen20153d} \cite{ibayashi2015dollhouse}\\ 
AR annotations & \cite{sodhi2013bethere} \cite{gurevich2012teleadvisor, gurevich2015design} \cite{wang2009experimental} \cite{gupta2016you} \cite{kim2018effect} \cite{wang2019head} \cite{norman2019impact} \cite{zillner2018augmented} \cite{higuch2016can} \cite{utzig2019augmented} \cite{nittala2015planwell} \cite{zenati2013new, zenati2015maintenance}\\   
Hands & \cite{wang2014mutual} \cite{sodhi2013bethere} \cite{sakong2006supporting} \cite{izadi2007c} \cite{gao2016oriented} \cite{lee2017sharedsphere, lee2018user} \cite{sasikumar2019wearable} \cite{alem2011handsonvideo} \cite{kim2019evaluating} \cite{higuch2016can} \cite{ibayashi2015dollhouse} \cite{sun2018employing, sun2016optobridge}\\
Audio Avatar& \cite{shen2010augmented, shen2006framework, shen2008product, shen2008collaborative} \cite{sidharta2006augmented} \cite{gauglitz2014world, gauglitz2012integrating, gauglitz2014touch} \cite{masai2016empathy, lee2016remote} \cite{tait2015effect} \cite{kurata2004remote} \cite{ou2003dove} \cite{lukosch2012novel} \cite{gronbaek2001interactive} \cite{pan2020posemmr} \cite{elvezio2017remote} \cite{luxenburger2016medicalvr}\\
\bottomrule
\end{tabulary}
\caption{Remote Collaboration Systems classified in avatar categories.}
\label{tab:avatar}
\end{table}

Avatars represent entities in virtual environments. 
We classify the most commonly used avatars in scientific literature as well as in commercial AR/VR/MR software in categories and differentiate them by descriptive terms (see Table \ref{tab:avatar}):
\begin{itemize}
\item \textbf{Realistic and Cartoon graphics}

We distinguish avatars through visualisation style, i.e. how the 3D model of the avatar is rendered (cartoon or realistic style). One of the reasons we use this as a descriptor is that we want to help researches who are focused on the appearance of avatars. Additionally, there is existing work which addresses certain research questions concerned with avatar visualisation styles. E.g. according to Yoon et al. \cite{yoon2019effect} there is no statistical difference in regards to social presence with different visualisation styles but that user perception differs between cartoon and realistic avatars.
A cartoon avatar allows for a more playful atmosphere, whereas realistic avatars tend to represent a professional environment.

\item \textbf{Avatar Type} 
We divide avatar types in subcategories: Full Body, Upper Body, Head \& Hands and Hands only. 
Full body avatar refers to humanoid avatars where all limbs are attached to it (e.g. hands, arms, legs etc.). 
An Upper Body avatar consists of a head, hands and torso but no legs. 
The Head \& Hands type of avatar is composed of a floating head combined with (detached) hands. Hands only means that a user is only represented by virtual hands.

\item \textbf{Reconstructed Model Avatar} 
If a system is capable of creating an avatar that resembles the respective user we categorise it as a system that uses \textit{Reconstructed Model} avatars. This category includes avatars that are created from face reconstruction in any form and excludes avatars that do not have a realistic face (e.g. reconstructed/personalized hands only is excluded). Non-reconstructed avatar means in general choosing from existing 3D models, without significant customization options.

\item \textbf{Video Avatar} 
Some systems implement avatars as video projections, similar to typical videoconferencing systems. Systems fall under this category if a user is seen as a video feed in an immersive virtual environment. Additionally, we categorise systems which use multiple cameras to reconstruct 3D video avatars here. As an example, the work of Matthes et al. \cite{matthes2019collaborative} implements such a system based on multiple depth cameras.

\item \textbf{Audio Avatar} 
Although avatars are often represented as a humanoid 3D model, the term avatar is in general used for any kind of user representation in virtual worlds, even including invisible forms. In this work, \textit{Audio Avatar} represents an entity in a system which enables communication with other users, regardless of visual appearance. Systems with this type do not rely on any visual form for users in remote collaboration systems. Users in such systems use audio for communicating with each other.

\item \textbf{AR Annotations} Systems which use no 3D model for other users but have annotations instead are in this category. This differs from \textit{Audio Avatar} in the sense that \textit{Audio Avatar} uses audio communication only, whereas in \textit{AR Annotations} the remote expert communicates with the local user with annotations, i.e. the other users presence is perceived through visual annotations. Unique avatars are usually not necessary in this scenario, because the roles are clearly separated and the users can distinguish each other by actions. In many systems using \textit{AR Annotation} avatars, audio communication is not implemented. As an example, the work of Zillner et al. \cite{zillner2018augmented} uses visual annotations such as text, pictures and freehand drawings to give precise instructions to the local worker without relying on audio communication.

\end{itemize}

A typical avatar configuration for VR based systems consists of the head (position of VR HMD) and hands (position of controllers) and is usually the most minimalistic avatar in VR scenarios. 
An avatar consisting solely of virtual hands tends to be used in AR based systems which use hand tracking or gesture techniques \cite{sodhi2013bethere, gao2016oriented}.
We identified a majority of \textit{Audio Avatars} in AR systems (see Table \ref{tab:avatar}).
While VR systems seem to always rely on 3D model representation of other users, AR based approaches often omit visual representation when using remote expert scenarios.
In such cases, the local and remote user communicate via audio with each other and share their view \cite{gauglitz2012integrating, gauglitz2014world, gauglitz2014touch}.
Furthermore, some systems rely mainly on visual annotations \cite{gurevich2015design, gurevich2015design, sodhi2013bethere}, which we marked as \textit{AR Annotations} in Table \ref{tab:avatar}.

Piumsomboon et al. \cite{piumsomboon2018mini} developed a system with an adaptive avatar \textit{Mini-Me} which uses redirected gaze and gestures to enhance remote collaboration with improved social presence.
To assess the usefulness of the avatar, a scenario where a remote expert in VR assists a local worker in AR was used.
The remote expert was shown to the AR user as a miniature avatar which was able to sucessfully transmit nonverbal communication cues according to the authors.
Although focusing on the novelty of the proposed avatar, the system proved to be useful for overall remote collaboration. \par

Elvezio et al. \cite{elvezio2017remote} developed a system with virtual twins of physical objects.
A remote expert uses virtual replicas of physically existing objects to guide a local user performing certain tasks with such objects.
In this case, communication with both users only takes place by transmitting the pose of the mentioned objects. 
In the work of Luxenburger et al. \cite{luxenburger2016medicalvr} the communication between users takes place through media sharing.
A user is filling out a report on a mobile device which is then visible to a remote user by means of a VR HMD.\par

In the commercial VR remote collaboration system EngageVR \cite{EngageVR}, users can create their own full body avatar with reconstructed face, by uploading a single picture. 
Machine learning techniques in the backend of the system reconstruct a fully textured 3D mesh of the head and attaches it automatically to a predefined body models.
Some other commercial systems are not as sophisticated and use cartoon like avatars \cite{Glue, breakroom, MeetInVR}.
Other popular systems such as VRChat \cite{chat2018create} or NeosVR \cite{NeosVR} allow users to create and upload their own avatars.
By means of an SDK, they can upload fully animated humanoid avatars regardless of their appearance.
The seamless integration of these arbitrary avatars is achieved by applying a specific skeletal structure to the model.

\subsection{Interaction}
\label{sec:interaction}
\begin{table*}[t]
\centering
\begin{tabulary}{0.95\textwidth}{p{13em}p{25em}}
\toprule
Interactive Feature& \hfil References \\
%\midrule
\cmidrule(rl){1-1}\cmidrule(rl){2-2}
Shared 3D Object Manipulation  & \cite{breakroom} \cite{EngageVR} \cite{Glue} \cite{MeetInVR} \cite{MozillaHubs} \cite{NvidiaHolodeck} \cite{Spatial} \cite{StageVR} \cite{TechViz} \cite{TheWild} \cite{ViveSync} \cite{WorldViz} \cite{regenbrecht2006carpeno} \cite{regenbrecht2004using} \cite{haller2005coeno} \cite{wang2014mutual} \cite{wang2009experimental} \cite{sakong2006supporting} \cite{sidharta2006augmented} \cite{shen2008product} \cite{shen2008collaborative} \cite{regenbrecht2002magicmeeting} \cite{orts2016holoportation} \cite{chat2018create} \cite{NeosVR} \cite{IrisVR} \cite{Acadicus} \cite{Rumii} \cite{Garou} \cite{MeetingRoom} \cite{BigScreen} \cite{VTime} \cite{SecondLife} \cite{pan2020posemmr} \cite{norman2019impact} \cite{tan2017virtual} \cite{galambos2015design, galambos2014merged} \cite{elvezio2017remote} \cite{lehner1997distributed} \cite{pouliquen2016remote} \cite{ibayashi2015dollhouse}\\
Media Sharing& \cite{breakroom} \cite{EngageVR} \cite{Glue} \cite{MeetInVR} \cite{MozillaHubs} \cite{NvidiaHolodeck}\cite{Spatial} \cite{StageVR} \cite{TechViz} \cite{TheWild} \cite{ViveSync} \cite{WorldViz} \cite{regenbrecht2006carpeno} \cite{regenbrecht2004using} \cite{haller2005coeno} \cite{wang2014mutual} \cite{izadi2007c} \cite{gronbaek2001interactive} \cite{speicher2018360anywhere} \cite{schafer2019towards} \cite{chat2018create} \cite{NeosVR} \cite{Rumii} \cite{VirBELA} \cite{Garou} \cite{MeetingRoom} \cite{BigScreen} \cite{Wave} \cite{VTime} \cite{SecondLife} \cite{tan2017virtual} \cite{galambos2015design, galambos2014merged} \cite{luxenburger2016medicalvr}\\

2D Drawing & \cite{EngageVR} \cite{Glue} \cite{MeetInVR} \cite{MozillaHubs} \cite{Spatial} \cite{StageVR} \cite{WorldViz} \cite{regenbrecht2006carpeno} \cite{haller2005coeno} \cite{gu2011technological} \cite{wang2014mutual} \cite{ou2003dove} \cite{izadi2007c} \cite{gronbaek2001interactive} \cite{speicher2018360anywhere} \cite{higuchi2015immerseboard} \cite{NeosVR} \cite{Rumii} \cite{MeetingRoom} \cite{galambos2015design, galambos2014merged}\\

Mid-Air Drawing in 3D& \cite{Glue} \cite{MeetInVR} \cite{MozillaHubs} \cite{Spatial} \cite{StageVR} \cite{WorldViz} \cite{NeosVR} \cite{IrisVR} \cite{zillner2018augmented} \\

Shared Gaze Awareness & \cite{poppe2011prototype, poppe2012preliminary} \cite{wang2014mutual} \cite{masai2016empathy} \cite{gupta2016you} \cite{piumsomboon2017empathic, piumsomboon2017covar} \cite{speicher2018360anywhere} \cite{higuchi2015immerseboard} \cite{wang2019head} \cite{norman2019impact} \cite{bai2020user} \cite{higuch2016can} \cite{utzig2019augmented}\\
Convey Facial Expression & \cite{masai2016empathy} \cite{lee2016remote} \cite{NeosVR} \cite{tan2017virtual}\\

Hand Gestures & \cite{Spatial} \cite{sodhi2013bethere} \cite{poppe2011prototype} \cite{wang2014mutual} \cite{wang2009experimental} \cite{sakong2006supporting} \cite{izadi2007c} \cite{lukosch2012novel} \cite{piumsomboon2017empathic, piumsomboon2017covar} \cite{gao2016oriented, gao2017static} \cite{lee2017sharedsphere, lee2018user} \cite{teo2019360drops} \cite{schafer2019towards} \cite{orts2016holoportation} \cite{NeosVR} \cite{sasikumar2019wearable} \cite{tan2017virtual} \cite{bai2020user} \cite{alem2011handsonvideo} \cite{kim2019evaluating} \cite{higuch2016can} \cite{ibayashi2015dollhouse} \cite{sun2018employing, sun2016optobridge}\\

AR annotations & \cite{Spatial} \cite{sodhi2013bethere} \cite{poppe2011prototype} \cite{poppe2012preliminary}  \cite{haller2005coeno} \cite{wang2014mutual} \cite{sidharta2006augmented} \cite{shen2010augmented, shen2008product,  shen2008collaborative} \cite{masai2016empathy, lee2016remote} \cite{tait2015effect} \cite{kurata2004remote} \cite{ou2003dove} \cite{lukosch2012novel} \cite{regenbrecht2002magicmeeting} \cite{speicher2018360anywhere} \cite{ryskeldiev2018spatial} \cite{lee2017sharedsphere, lee2018user} \cite{teo2019360drops} \cite{kim2018effect, kim2019evaluating} \cite{sasikumar2019wearable} \cite{alem2011handsonvideo} \cite{zillner2018augmented} \cite{higuch2016can} \cite{chen20153d}  \cite{nittala2015planwell} \cite{zenati2013new, zenati2015maintenance}\\

AR viewport sharing & \cite{tait2015effect} \cite{kurata2004remote} \cite{piumsomboon2017empathic, piumsomboon2017covar} \cite{gao2016oriented, gao2017static} \cite{speicher2018360anywhere} \cite{ryskeldiev2018spatial, ryskeldiev2018streamspace} \cite{lee2017sharedsphere, lee2018user} \cite{teo2019360drops, teo2019merging} \cite{kim2018effect}  \cite{sasikumar2019wearable}  \cite{nittala2015planwell} \cite{sun2018employing, sun2016optobridge} \cite{zenati2013new, zenati2015maintenance}\\

\bottomrule
\end{tabulary}
\caption{Shared interaction possibilities among remote collaboration systems.}
\label{tab:interaction}
\end{table*}

We identified common interactive elements in remote collaboration systems that are found in various works and literature. 
In Table \ref{tab:interaction} we provide an overview of literature and work which is categorised in multiple different interaction categories. It is to note that a category is not mutually exclusive to another, e.g. a system which uses media sharing might also use hand gestures. This section explains each interaction technique with a few examples.
The common features we extracted are as follows:
\begin{enumerate}
    \item Shared 3D Object Manipulation
    \item Media Sharing
    \item AR Annotations
    \item 2D Drawing
    \item AR Viewport Sharing
    \item Mid-Air Drawing in 3D
    \item Hand Gestures
    \item Shared Gaze Awareness
    \item Convey Facial Expression
\end{enumerate}
Table \ref{tab:interaction} guides the reader to interesting and major publications which use the mentioned interaction techniques. 
\subsubsection{Shared 3D object Manipulation}
The most commonly shared feature between remote collaboration systems is the possibility to interact and manipulate shared 3D objects in a virtual space. 
The type of interaction differs between systems, but the focus is on manipulating one or many 3D objects.
AR technology is used by Shen et al. \cite{shen2008collaborative, shen2008product, shen2010augmented} where multiple users interact with 3D objects in a collaborative AR environment. 
A stylus with two markers attached is used as additional interaction tool which enables feature highlighting and 3d object manipulation.
More recent work uses hand tracking/gestures to interact with objects \cite{Spatial}. Schlünsen et al. \cite{schlunsen2019vr} compared free-hand-manipulation with widget-based manipulation techniques.
Their study shows that free-hand interaction is preferred over widget-based interaction by users. \par

\subsubsection{Media Sharing}
Systems which are able to share documents, images, videos and other form of media are categorised here.
Haller et al. \cite{haller2005coeno} created a system with a tangible interface, a table with touchscreen for media sharing.
It additionally featured sharing media from desktop applications to the tabletop.
A web-based VR solution was developed by Monahan et al. \cite{monahan2008virtual} which implements media sharing such as videos and images in an educational context. 

\subsubsection{AR Annotations}
One of the most commonly used tool for communication and interaction in AR based systems is annotations.
Annotation types include 2D drawing, text, or simple pointers.
The work of Speicher et al. \cite{speicher2018360anywhere} utilizes a 360 camera to capture the surroundings of one user, while other users are able to draw and annotate on the input stream.
The annotations and drawings are then visualised by a projector to the local user's physical space.
Kurata et al. \cite{kurata2004remote} present a wearable HMD which receives remotely annotated input in form of drawings.
A special feature of this system is a laser pointer that enables the wearer of an HMD to draw the attention of remote users towards a certain object by pointing on it.

\subsubsection{2D Drawing}

Especially in remote collaboration systems with focus on replicating a virtual meeting scenario, drawing on surfaces is a widely used feature.
In more sophisticated systems such as Glue Collaboration Platform \cite{Glue}, users are able to place a virtual whiteboard.
This whiteboard can be re-positioned and resized allowing multiple users to draw with virtual pens in many sizes and colors. Mimicking real world objects, it is also possible to use an eraser.

\subsubsection{AR viewport sharing}
This category includes systems which implemented sharing a user's view perspective.
Tait et al. \cite{tait2015effect} created a system which reconstructs a local user's environment by using depth sensors attached to an HMD. 
Reconstructing the environment from a local scene, remote users can move independently through the virtual environment. 
The local user is then represented as a frustum in the reconstructed scene, allowing the remote user to see where the local counterpart is looking at. 
The study of Tait et al. \cite{tait2015effect} suggests that implementing view independence between local and remote user improves task completion time. Sasikumar et al. \cite{sasikumar2019wearable} combined AR and VR users together and enabled view frustum sharing which is visible to the local user as a grey cuboid. The goal of their work was to convey nonverbal communication cues such as eye gaze and hand gestures.

\subsubsection{Mid-Air Drawing in 3D}
Mid-Air drawing allows users to create 3D paintings, which can then be observed from multiple users in different angles.
This interaction method seems to be mostly available in commercial systems such as Glue Collab \cite{Glue} or \cite{MozillaHubs}. A user is able to draw in the air of the virtual world by utilizing a VR controller as virtual pen.
Other systems such as the one proposed by Zillner et al. \cite{zillner2018augmented} implement a remote expert scenario, where one user is streaming his surroundings with an RGB-D camera and a remote expert is observing and annotating for assistance.
The remote expert is able to segment objects, to create animations, to draw on geometry and to place annotations which can be viewed by the AR user.
\subsubsection{Hand Gestures}
Systems which utilize hand gestures and convey hand movements through the remote collaborative space are included in this category.
Sophistated systems such as \textit{Spatial} \cite{Spatial} use AR HMD's to enable a full interaction with the 3D environment via a hand tracker.
Tan et al. \cite{tan2017virtual} implemented a VR telepresence system which allowed multiple users to interact with objects and watch videos together. The authors used motion capture gloves to animate arms, hands and fingers of a VR avatar.
Kim et al. \cite{kim2019evaluating} implemented a MR collaboration system to evaluate combinations of visual communication cues using gestures.
The authors found that certain combinations of communication cues such as \textit{hand visualisation} together with \textit{finger pointing direction} does not provide any significant benefit for remote collaboration.

\subsubsection{Shared Gaze Awareness}
Systems which allow users to share gaze awareness belong into this category.
We include systems that allow precise tracking and transmission of gaze awareness and exclude systems that indicate gaze perception by head rotation only. E.g. Galombos et al. \cite{galambos2015design} is not included since the gaze directon of a user is only indicated by the direction the avatar is facing.
As an example, Speicher et al. \cite{speicher2018360anywhere} created a system that allows the participants to show exactly which position they are looking at in a 360° video feed. 
The work of Poppe et al. \cite{poppe2011prototype, poppe2012preliminary} uses avatars around a virtually augmented table and positions them according to the gaze information of the corresponding user.
Billinghurst et al. developed \textit{Emapthy Glasses} \cite{masai2016empathy, lee2016remote} which is a HMD that enables streaming a live video feed with accurate gaze information.
Norman et al. \cite{norman2019impact} implemented a system which gives direct visual feedback of other user's gaze behavior. 
Using a system that combines multiple AR HMD's with a desktop PC, participants are asked to place virtual furniture on a regular table in a collaborative manner.

\subsubsection{Convey Facial Expression}
This category addresses work which is able to transmit facial expressions to other users. Systems which use video transmission of other user's faces are not included.
Lee \cite{lee2016remote} and Masai\cite{masai2016empathy} et al. use \textit{Empathy Glasses} to transmit facial expressions bidirectionally. 
A local user's face is analyzed by the built in modules of the glasses, while a remote user's expression is tracked via webcam.
The authors of \cite{tan2017virtual} integrated lip syncing into their VR remote telepresence system in order to enable a more immersive communication experience.

\section{Professional meetings through virtual and augmented remote collaboration}

Currently, there are already many tools and collaboration systems available which utilize virtual or augmented reality. In this section we consider commercial and professional systems which support more than 10 users simultaneously \cite{breakroom, EngageVR, Glue, MeetInVR, MozillaHubs, NvidiaHolodeck, Spatial, StageVR, TechViz, TheWild, WorldViz}. 
During the COVID-19 outbreak in early 2020, one of the biggest scientific conferences for virtual reality research was held completely online and virtual. 
To this date, it was the first major conference held completely virtually.
During this time, the organizing committee faced a very challenging task to provide a pleasant conference experience for all participants without the need for a physical presence.
During this precedent case, the whole conference featured livestreams for each track, where each presenter had the opportunity to either give a presentation with pre-recorded video, live video transmission or presenting with a VR HMD.
Additionally, there where multiple virtual meeting rooms in which participants could join and then interact and network with other users using VR HMD's.
The virtual meeting rooms utilized during this conference where built on  Mozilla Hubs \cite{MozillaHubs}.
It features a web-based meeting room creation software, which enables users to develop and maintain their own meeting experience.
Utilizing a cartoon graphics style, this system works with a desktop application, internet browser and even mobile devices, covering a broad possible user audience.
Avatars are chosen from existing, pre-defined models.
They use a mix between an abstract representation and Upper Body avatars by using a robot-like representation for users.\par

While this solution uses an open source approach, there are several commercial products available \cite{Glue, MeetInVR, TheWild, StageVR, breakroom} which are similar in terms of remote collaboration utlizing VR technology.
\textit{Glue Platform} \cite{Glue} is a platform for business professionals offering immersive 3D graphics in cartoon style.
It is built to be used with virtual reality HMD's and claims to be an extension to the everyday working life.
Main features include spatial audio, 3D avatars, interactive and persistent objects.
The avatars use a simple Head \& Hands approach.

Another commercial software available is called Breakroom \cite{breakroom}.
It supports VR HMD's, is available for multiple platforms, and features full body avatars.

\subsection{Comparison of commercial systems}
In this section, we compare existing commercial/professional remote collaboration systems which are already in use by industry and individuals. 
We found several VR remote collaboration systems which are used in a professional context \cite{NvidiaHolodeck, Glue, Spatial, EngageVR}.
These systems have many common aspects, such as allowing many users to participate, or the collaboration tools available in  the virtual environment. Commercial systems seem to differ mostly in terms of avatars, environments and visualisation styles. 
Especially in interaction, the systems share many common collaboration tools inside the virtual world.
A typical way to collaborate in these systems is using some drawing mechanism  or mid-air drawing in 3D space.
Copying real world interaction methods, users can sketch on virtual black- or whiteboards and share their results in real time with each other.
Some of the mentioned systems also implement mid-air drawing, and allow sketches to be observed from multiple angles. 
Usually there is also a name indicator, displaying the name of participants.
This seems to be necessary even in systems with completely reconstructed or personalised avatars such as EngageVR \cite{EngageVR} and Spatial \cite{Spatial}.
A common use case in the aforementioned systems is sharing and observing virtual objects together.
Systems such as EngageVR \cite{EngageVR} or Neos VR \cite{NeosVR} allow users to place any 3D object previously added to a catalogue.
In some cases it is also possible to show other users information by floating markers i.e. annotations.
The virtual environment in these systems usually have a table and multiple seats to copy the physical space of real world meetings.
A common scenario involves users to sit on virtual chairs and to present on a virtual tv or projector.
Some more advanced systems (e.g. Mozilla Hubs \cite{MozillaHubs, NeosVR}) allow screen sharing to the virtual world.
Even more sophisticated systems (e.g. EngageVR \cite{EngageVR}) extends screen sharing functionality with full control over normal desktop applications in the virtual world.
Additionally, most developers are implementing platform independence, supporting devices such as desktop, VR HMD's, tablets and smartphones.
Mozilla Hubs \cite{MozillaHubs} is fully available in a web browser.

\subsection{Exploring the strengths and weaknesses of commercial remote collaboration systems}
As part of this survey, we identified strengths and weaknesses of popular commercial virtual meeting systems. This survey includes a total of 25 commercial and professional systems (24 VR and one MR based system).
The combined strength of these systems include:
\begin{enumerate}
    \item Many users are able to join and participate in virtual meetings simultaneously \cite{MozillaHubs, breakroom, MeetInVR}, usually allowing about 20-50 people to share a virtual space simultaneously
    \item Intuitive interaction possibilities such as mid-air drawing \cite{MozillaHubs}, drawing on white- or blackboards \cite{MeetInVR}, media sharing and screen sharing \cite{EngageVR}
    \item Spatial audio which enables localisation of an audio source during meetings more naturally \cite{Glue}
    \item Persistent virtual objects which exist through multiple sessions (e.g. a drawing from a session before is still present in the next session) \cite{Glue}
    \item Placing arbitrary 3D objects in a shared virtual environment \cite{EngageVR, NeosVR}
    \item Reconstructed, personalised avatars \cite{EngageVR} and user created avatars through an API which is provided by the developers \cite{chat2018create, NeosVR}
    \item Availability on multiple platforms: desktop, mobile devices and web browser \cite{MozillaHubs, Glue}
\end{enumerate}
Some systems are enterprise solutions \cite{Glue, NvidiaHolodeck}, tailored to the specific needs of companies, what renders them unattractive or even inaccessible to the general public.
One major issue with these systems is the missing transmission of nonverbal communication cues to the virtual environment by means of an avatar, which is an important feature of traditional face-to-face collaboration.

Another weak point of the current systems is dynamic content creation for the virtual environment.
These systems are limited to the choice of virtual environments provided by the developer or need expert knowledge to create them \cite{MozillaHubs, EngageVR, MeetInVR, Glue}. \par
Some systems implement a whole Metaverse that acts as a virtual space for many scenarios such as training, collaboration and other social activities inside a virtual world. NeosVR \cite{NeosVR} and VRChat \cite{chat2018create} allow experienced users to create custom environments, avatars and interactive objects by providing an API within the game engine in which the platform is implemented. \par
Many systems excel in certain aspects but lack novelty in others. As an example, the immersive remote collaboration software VirBELA \cite{VirBELA} allows hundreds of users to participate in a virtual world simultaneously but the avatars lack personalisation.

\section{Discussion and Survey Result}
In this section we provide an overview of the insights and statistical data we gathered throughout the survey.
Especially, we want to emphasize on our taxonomy consisting of \textit{Environment}, \textit{Avatars} and \textit{Interaction} and important findings in each category.
It is to note that we included professional and commercial systems in our survey (24 VR based systems and one MR based system).
This implies that some of the discussed applications are not published in scientific articles and the implementation of interactive features, virtual environment and the audiovisual representation of users is subject to change in the future. E.g. a discussed system does not implement full body avatars at the time of writing, but could implement it later on. 

\subsection{Environment}
We categorised remote collaboration systems according to their technology: AR/VR/MR, use case, visualisation styles and sensory inputs. The technology distribution of included systems is shown in Figure \ref{fig:technologydistributionchart}.

\subsubsection{Technology}
In the category of VR based synchronous remote collaboration systems we included 24 commercial and 7 research oriented systems.
The majority of systems with VR technology are commercial systems, which could be an indicator that VR based systems are currently more under development in the industrial sector rather than the research community and therefore could be placed on the plateau of productivity. 
17 purely AR based systems where found in which users are able to communicate and collaborate in real-time by means of Video-Seethrough, Optical-Seethrough AR or a combination of both.
To the best of our knowledge there is no commercial system which is solely based on AR allowing real-time synchronous remote collaboration.
MR based systems form the majority of the discussed systems. 33 research oriented and one commercial system are included in this category.
\begin{figure}[h]
\centering
\includegraphics[width=0.7\columnwidth]{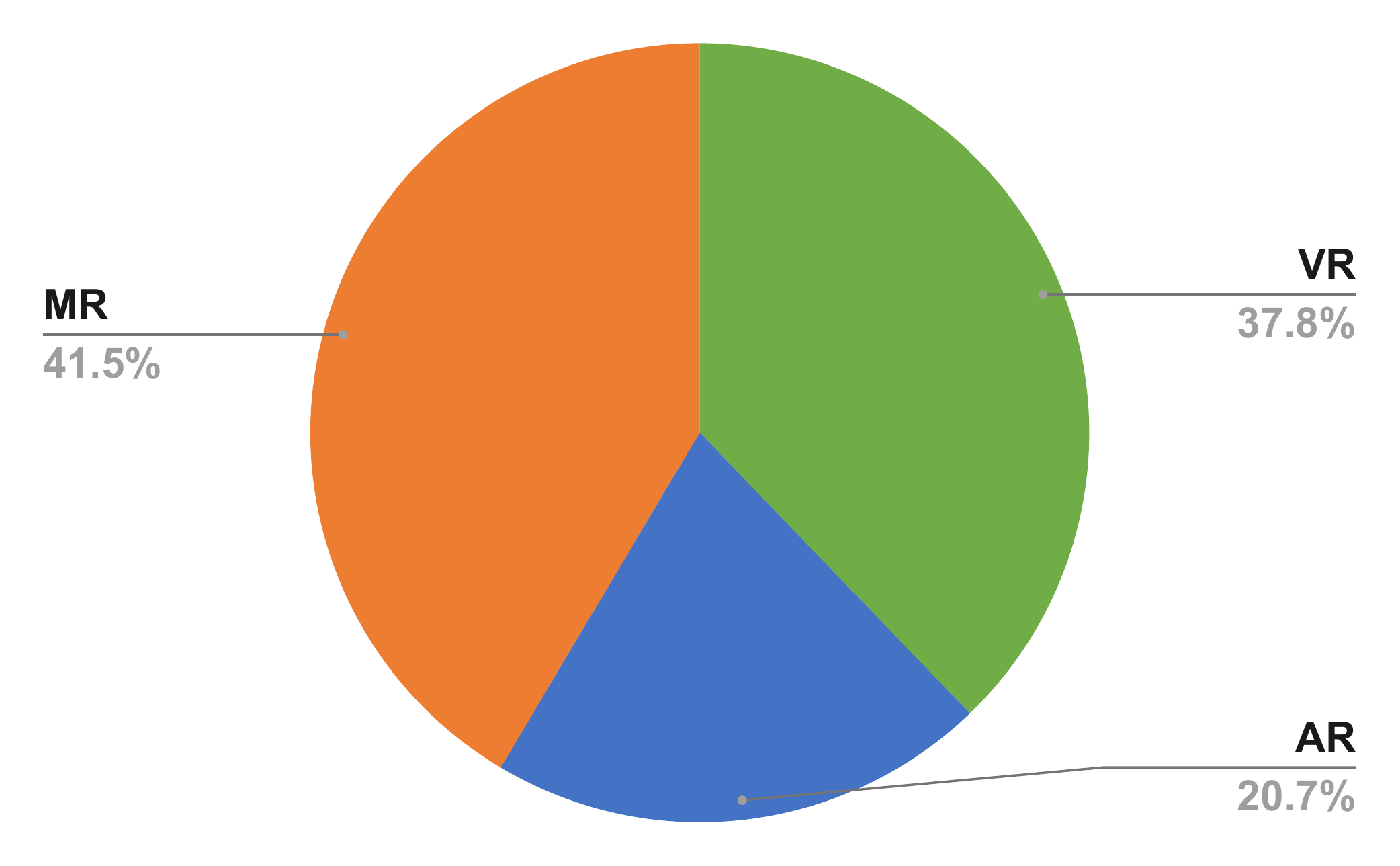}
\caption{Distribution of used technology in the discussed systems.}
\label{fig:technologydistributionchart}
\end{figure}

\subsubsection{Use cases}
We divided systems into three different main use cases: \textit{Meeting, Design} and \textit{Remote Expert} as they are most popular in the literature. 
The distribution of mentioned use cases with respect to their technology is shown as a graph in Figure \ref{fig:systemsforeachusecase}.

Systems based on VR technlogy often involve many users (more than two), with a focus on Meeting and Design use cases.
The virtual environments used in VR technology tend to involve all participants equally, i.e. they can see the same things and have the same input modalities, whereas MR based systems often have asymmetric inputs.
For example, an asymmetric input method would be when two users share a virtual environment and one has a keyboard and the other has a VR headset with controllers as input device.
There was no purely VR based system which implemented a \textit{Remote Expert} scenario although it is by far the most popular use case in AR based systems.
While VR systems have a focus on \textit{Meeting} scenarios and AR systems a focus on \textit{Remote Expert} scenarios, MR systems are more distributed throughout use cases.
This indicates a strong correlation between the hardware and its respective benefits in certain use cases. For example, VR HMD's are more beneficial in meeting scenarios with an immersive, shared virtual environment, while AR HMDs have more advantages in supporting a user with a remote expert.
Since MR is a mix between AR and VR, the distribution of cases is more equally.
\begin{figure}[H]
\centering
\includegraphics[width=1.0\columnwidth]{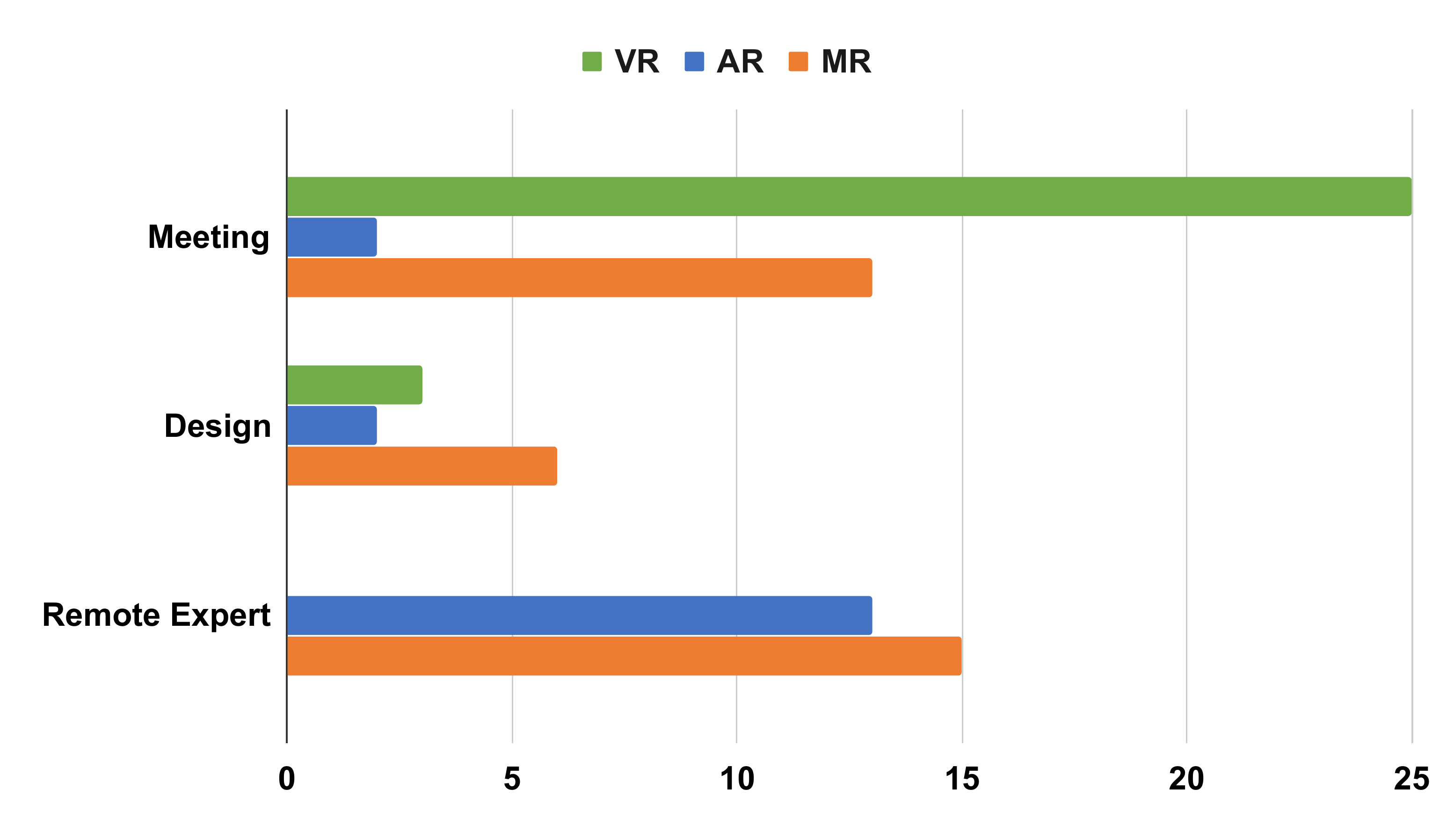}
\caption{Distribution of use cases in the discussed systems.}
\label{fig:systemsforeachusecase}
\end{figure}
\subsubsection{Visualisation Styles}
We did not find any hints that there is a clear deviation in respect to visualisation styles of synchronous remote collaboration systems. We differentiated between cartoon, realistic and annotation style. A system was labeled cartoon style when there was an obvious simplification of 3D objects with stylized visualisation. Older systems with realistic visualisation are also counted here, as long as no stylized visualisation technique was used. Many AR based systems use annotated video only while some use rendered 3D objects with a cartoon style. 
\subsubsection{Sensory Impressions}
The recent literature seems to focus more on audiovisual systems and does not support other sensory impressions such as olfactory and tactile cues.
Although tangible interfaces where popular especially in AR technology, the focus is drifting towards audiovisual systems.
One of the reasons might be that markers are no longer required to be placed on physical objects which often already implied a tangible system, if markers are placed on non-stationary objects. Additionally, the tracking accuracy of AR systems is constantly being improved. An example is Rambach et al. \cite{rambach2017poster, rambach2019slamcraft}, where the authors use SLAM technology for accurate object tracking without markers.
\iffalse
\begin{figure}[H]
\centering
\includegraphics[width=1.0\columnwidth]{Images/TechnologyDistributionChart.pdf}
\caption{Distribution of used technology in the discussed systems.}
\label{fig:technologydistributionchart}
\end{figure}
\subsection{Interaction}
\begin{figure}[ht]
\centering
\includegraphics[width=1.0\columnwidth]{Images/systemsforeachusecase.pdf}
\caption{Distribution of use cases in the discussed systems.}
\label{fig:systemsforeachusecase}
\end{figure}
\fi

\subsection{Interaction}

\iffalse
\begin{figure}[ht]
\centering
\includegraphics[width=1.0\columnwidth]{Images/InteractionTypesGraph.pdf}
\caption{Interaction Types sorted by the number of implementations in the presented systems.}
\label{fig:interactiontypesgraph}
\end{figure}
\fi

Examining the different types of interaction in remote collaboration systems, a general majority of the interaction type \textit{Shared Object Manipulation} is found. A possible explanation for this is the ease of implementation and general versatility of a task involving the manipulation of 3D objects. Another popular feature is \textit{Media Sharing}, i.e. possibility to share images, videos and other form of media. Some interaction types are more popular in VR technology, such as \textit{Shared Object Manipulation, Media Sharing, 2D Drawing} and \textit{Mid-Air Drawing in 3D}. The interaction types \textit{Viewport Sharing, AR annotations} and \textit{Shared Gaze Awareness} had no implementations at all in a pure VR scenario. Transmitting facial expressions by using avatars is by far the least prominent feature in remote collaboration systems, even though it is an important step towards more natural conversations over distance. A comparative graph about the interaction types and their presence in the discussed work is shown in Figure \ref{fig:interactiontypesgraph}.
\begin{figure}[H]
\centering
\includegraphics[width=1.0\columnwidth]{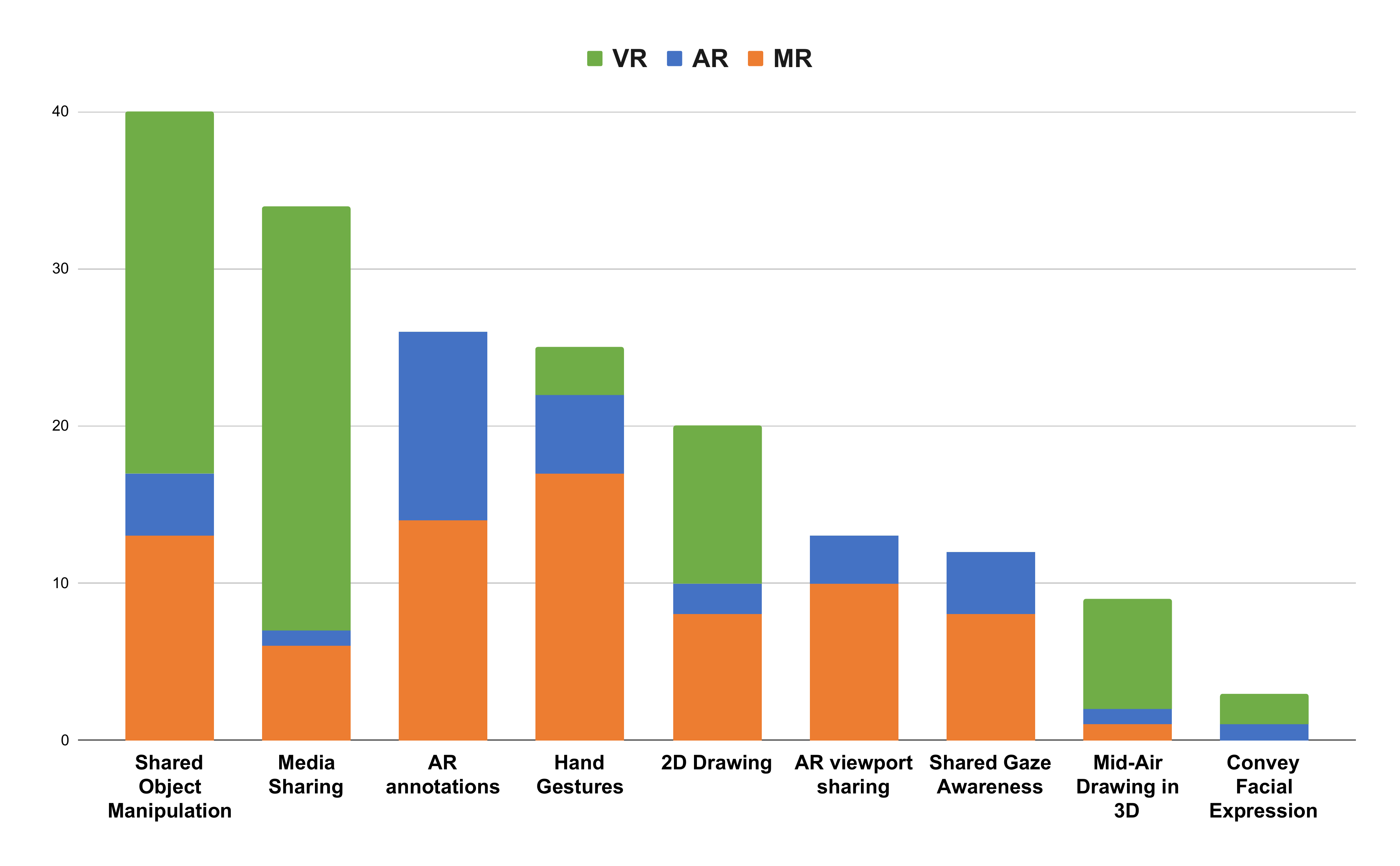}
\caption{Interaction Types in respect to the used technology.}
\label{fig:interactiontypesgraph}
\end{figure}

Additionally, we found that most of the commercial remote collaboration systems rely solely on VR technology (24 VR systems and one MR system).
Since the purchase of commercially available VR HMD's and a reasonable PC is required to operate these systems, their actual use is still limited. 
Therefore most professional and commercial systems tend to implement a desktop and mobile version of the VR application.
Only a few companies, such as Spatial \cite{Spatial} focus on integrating AR, VR, desktop and mobile together.
Deploying the same application on multiple hardware devices, each with different input modalities, raises new issues such as asymmetric input.
For example, a mobile device user can most likely collaborate only by voice or minimal interaction within a virtual environment, compared to a user with a VR HMD and full body tracking.
In an attempt to solve the asymmetric input problem, Fleury et al. \cite{fleury2015remote} investigated how the same interaction possibilities can be realized in virtual space, in a CAVE environment, and a high-resolution 2D display in wall format. 
Another approach is taken by Pouliquen-Lardy et al. \cite{pouliquen2016remote}, which implemented a multi-user remote collaboration MR system to study the asymmetric effects of different input modalities.
The authors used an approach with two different roles, a guide who could observe and communicate via audio, and a manipulator who was able to manipulate a 3D object.
The overall result of their study suggests that it is not necessary to develop symmetric interaction for all users during remote collaboration, but rather the same interaction possibilities for roles. For example, a user in the role of a guide should be able to observe and communicate via audio regardless of the hardware they are using.

\subsection{Avatars}
\begin{figure}[H]
\centering
\includegraphics[width=0.7\columnwidth]{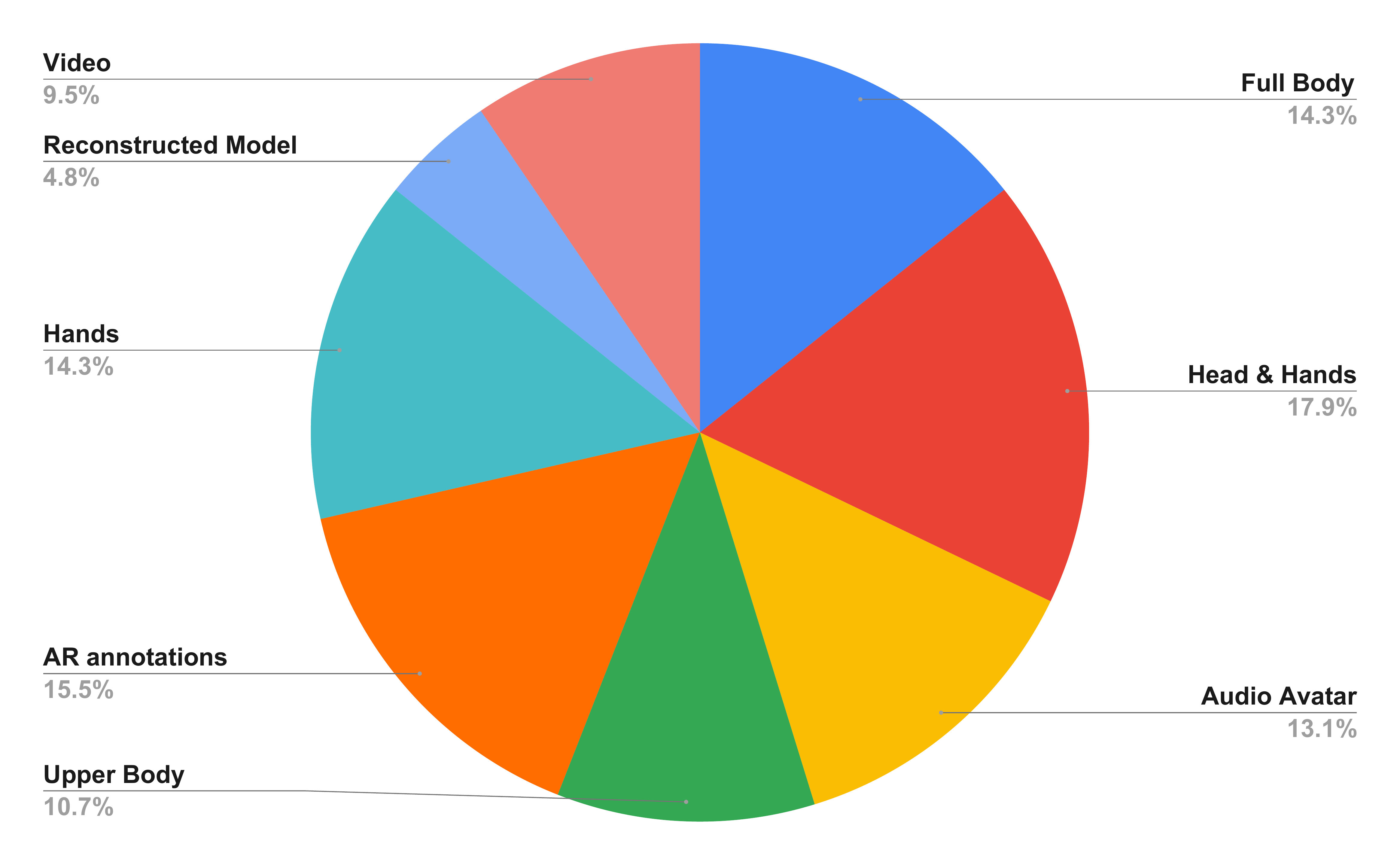}
\caption{Different avatar types in presented sytems.}
\label{fig:avatartypeschart}
\end{figure}

\begin{figure}[H]
\centering
\includegraphics[width=1.0\columnwidth]{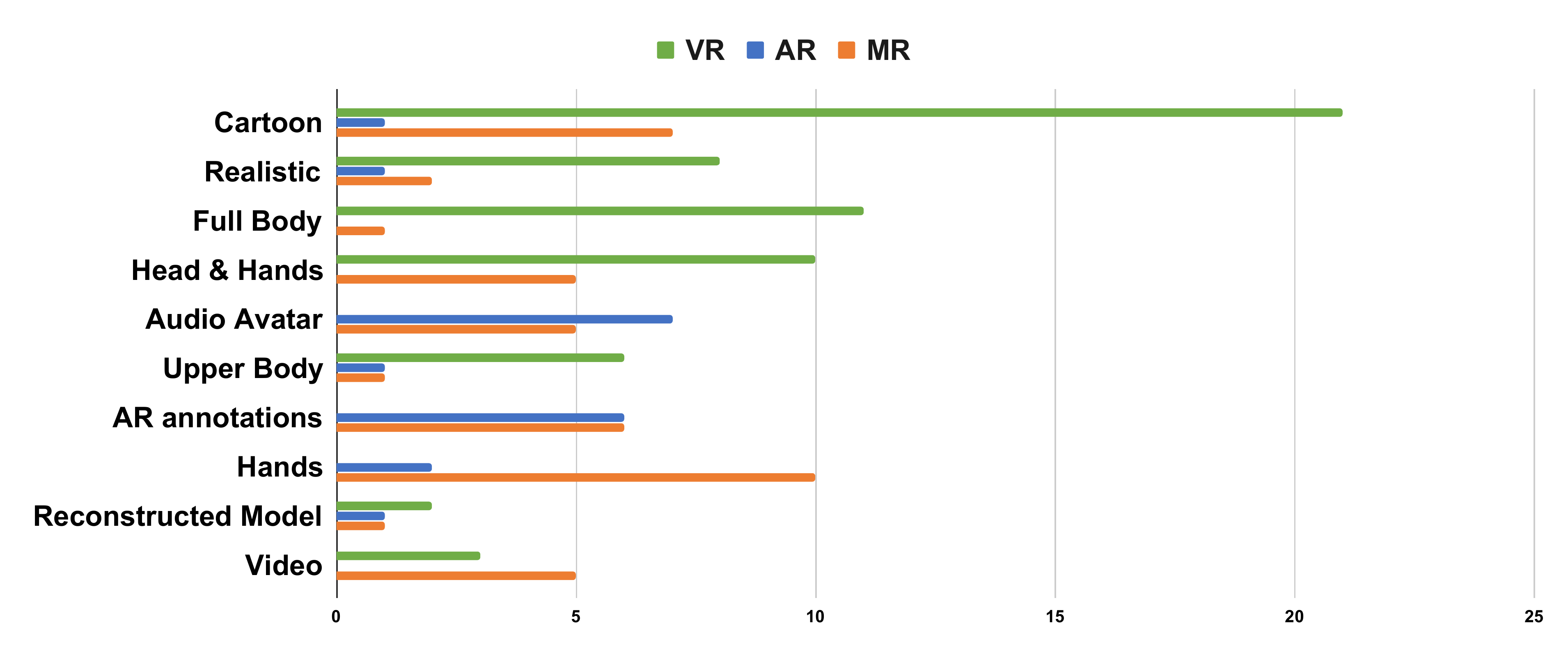}
\caption{Different avatar types in respect to their used technology.}
\label{fig:avatartechonologydistribution}
\end{figure}

The systems were analysed with respect to their specific avatar implementation. An explanation of our avatar categorisation is found in section \ref{sec:avatars}. We found that there is not a single most used avatar type for AR, VR or MR systems. However, we have seen that certain types of avatars are not used in combination with particular technologies. An overall distribution of avatars in the discussed systems is shown in Figure \ref{fig:avatartypeschart}.
Filtering the avatar types by technology gives interesting insights into the spectrum of personal embodiment in synchronous remote collaboration systems.
Looking at Figure \ref{fig:avatartechonologydistribution} we found that \textit{Full Body, Head \& Hands} and \textit{Upper Body} are most prominent in VR based systems.
To the best of our knowledge, there is no AR system that uses a full body 3D model for representation of other users.
The closest approach to a full body 3D model representation in an AR system is Holoportation \cite{orts2016holoportation}, but since several RGB-Depth cameras are involved that transmit video in real-time, we classify this approach as a video-based avatar system.
The \textit{Hands} approach for avatars is mostly used in AR and MR systems.
This is due to the fact that hand gestures provide a natural and easyily understood way to convey visual communication cues in a \textit{Remote Expert} scenario, which is used in AR and MR based systems. 
Overall we found a lack of systems which utilizes \textit{Reconstructed Models}, i.e. an avatar that is created through face reconstruction in the virtual environment. A reason for this is that the technology for easy reconstruction of humans is not yet widely available for researchers and industry implementing such systems.
Not surprisingly, \textit{Audio Avatars} are mostly used in AR and MR based systems. 
One of the reasons is that the shared virtual environment is formed through video transmission and the communication with other users happens through audio. 
VR systems use a rendered virtual 3D scene that simplifies finding other users in a virtual environment, allowing for a more sophisticated visualisation approach such as \textit{Full Body} or other types of visual avatars.

\section{Conclusion}
During this survey, we identified shared interactive elements, various types of avatars, and different kinds of virtual environments for synchronous remote collaboration systems.
To help researchers in various research fields we created a taxonomy consisting of \textit{Environment, Interaction} and \textit{Avatars}.
This approach aims to provide condensed information to specific topics which needs to be addressed while designing and implementing a synchronous remote collaboration system.
It appears that AR based systems focus on sharing a person's surroundings with emphasis on stimulating the audiovisual senses. 
This has recently led to a focus on remote and local user scenarios in AR/MR systems.
In such scenarios, typically a user is physically present at a certain location and shares his/her environment, which is then perceived by other users as a virtual world that can be augmented with virtual objects.
This means however, that each type of user (local user or remote expert) has different input and output options in their respective virtual environment. 
As an example, a local user streams the surroundings with an AR HMD, while a remote user can add annotations by using e.g. a VR HMD or tablet device, which defines separate roles during remote collaboration. \par
In contrast, VR based systems tend to involve all participants equally, allowing each user to participate in the same virtual environment and the same communication and collaboration possibilities. 
We also noticed a trend in VR technology to focus more on design and meetings rather than remote expert scenarios, as is mainly the case with AR and MR systems.
Independently between AR, VR and MR the research focus in remote collaboration software drifts towards integrating non-verbal communication cues.
Some researchers focus on developing solutions for nonverbal communication transmission, but this has not yet been integrated in professional and commercial remote collaboration systems.
This research includes Wang et al. \cite{wang2014mutual} focusing on eye-contact, Masai et al. \cite{masai2016empathy} conveying facial expressions and Lei et al. \cite{gao2017real} implementing mutual hand gestures to name a few.
\begin{acks}
This work was partially funded by Offene Digitalisierungsallianz Pfalz, BMBF, grant number 03IHS075B. This work was also supperted by the EU Research and Innovation programme Horizon 2020 under the grant agreement ID: 883293.
\end{acks}

\bibliographystyle{ACM-Reference-Format}
\bibliography{sample-base}

\end{document}